\documentclass[twocolumn,aps,superscriptaddress,showpacs, showkeys, nofootinbib,floatfix,linenumbers]{revtex4}
\usepackage{epsfig,bm,feynmf}
\usepackage{graphics}
\usepackage{amsmath}
\usepackage{mathrsfs}
\usepackage[normalem]{ulem}  
\usepackage[dvips]{color} 

\renewcommand{\sout}{\bgroup \color{red} \ULdepth=-.5ex \ULset}

\newcommand{\ket}[1]{\left| #1 \right\rangle}

\newcommand{\average}[1]{\left\langle #1 \right\rangle}

\begin{document}

\preprint{APS/123-QED}
\title{Spin Asymmetry of $J/\psi$ in Peripheral Pb+Pb Collisions at LHC}

\author{Yunpeng Liu}
\email{yliu@comp.tamu.edu}
\affiliation{Cyclotron Institute, Texas A$\&$M University, College Station, Texas 77843, USA}

\author{Carsten Greiner}
\email{Carsten.Greiner@th.physik.uni-frankfurt.de}
\affiliation{Institut f\"ur Theoretische Physik, Johann Wolfgang Goethe-Universit\"at Frankfurt, Max-von-Laue-Str. 1, D-60438 Frankfurt am Main, Germany}

 \author{Che Ming Ko}
 \email{ko@comp.tamu.edu}
\affiliation{Cyclotron Institute, Texas A$\&$M University, College Station, Texas 77843, USA}%
\affiliation{Department of Physics and Astronomy, Texas A$\&$M University, College Station, Texas 77843, USA}%

\date{\today}

\begin{abstract}
   By generalizing the statistical model for particle production to the spin degree of freedom of  initially produced $J/\psi$, we study the spin projection $J_y$ of $J/\psi$ perpendicular to the reaction plane in peripheral  heavy ion collisions at the LHC energy that leads to a strong, albeit of short duration, magnetic field. We find that for $J/\psi$s produced directly from charm and anticharm quarks in the color singlet state, like that in the Color-Singlet Model, their  yield  in the presence of the magnetic field is larger for $J_y=0$  than  for $J_y=1$ or $-1$. This leads to a spin asymmetry of finally produced $J/\psi$ even after including their final-state scattering in the produced quark-gluon plasma.   
\end{abstract}



\maketitle


\section{Introduction}

Compared with light hadrons, the $J/\psi$\ has larger binding energy and was suggested to survive high temperature 
and to be used as a probe of the early stage of the  hot dense matter produced in relativistic heavy ion collisions~\cite{Matsui:1986dk}. Many observables on $J/\psi$, including the yield, the transverse momentum distribution, the rapidity distribution, and the elliptic flow, have been measured at the Super Proton Synchrotron (SPS)~\cite{Gonin:1996wn, Abreu:1999qw, Abreu:2000xe}, the Relativistic Heavy Ion Collider (RHIC)~\cite{Adler:2005ph, Adare:2006ns, Abelev:2009qaa, Adamczyk:2012ey, Reed:2011zza, Qiu:2012zz, Adare:2008sh, Adare:2011yf, Adamczyk:2012pw}, and  the Large Hadron Collider (LHC)~\cite{Abelev:2012rv, Chatrchyan:2012np, Massacrier:2012yj}, and they have also been studied using  various  models~\cite{Matsui:1986dk, Xu:1995eb, Brambilla:2004jw, Brezinski:2011ju, Grandchamp:2001pf, Zhao:2007hh, Song:2011xi, Song:2013lov, Zhu:2004nw, Yan:2006ve, Marasinghe:2011bt}. All these observables are related to the identification of $J/\psi$s and the measurement of their momenta.  Although the spin degree of freedom of $J/\psi$ has recently also attracted much attention in p+p collisions~\cite{Chao:2012iv, Lansberg:2010vq, Adare:2010bd, Abelev:2011md, Adamczyk:x0}, there is little discussion on the spin of $J/\psi$ in A+A collisions. Because a strong magnetic field is produced transiently by the fast moving charged ions in their peripheral collisions~\cite{Rafelski:1975rf}, which can reach an energy scale of GeV if the collision energy is very high~\cite{Deng:2012pc}, it raises the interesting question on whether one can use  it to probe how the $J/\psi$ is produced from initial hard  collisions.

In this paper, we calculate the spin asymmetry of prompt $J/\psi$ at midrapidity in peripheral Pb+Pb collisions at the LHC energy by taking into account the effect of the magnetic field on its constituent heavy quarks that are produced from the fusion of gluons in initial binary collisions. Because charm and anticharm quarks carry opposite charges, their total magnetic moment is zero when their spins are parallel and nonzero when their spins are antiparallel.  The energies of $J/\psi$ with spin $J_y=0$ and  with spin $J_y=\pm1$ in the direction of the magnetic field are thus different, and this is expected to lead to different yields for $J/\psi$s with $J_y=0$ and  $J_y=\pm 1$ or a spin asymmetry of $J/\psi$. Because the strong magnetic field only lasts a short time in heavy ion collisions, its effect on the $J/\psi$ spin asymmetry depends on how $J/\psi$ is produced in the initial hard collisions. In the  Color-Singlet Model (CSM) for $J/\psi$ production~\cite{Chang:1979nn},  the $J/\psi$ is formed directly  from the gluon fusion when the strength of the external magnetic field may still be appreciable. The spin state of $J/\psi$ in this case is the same as that of the charm and anticharm quarks, and the relative yield of $J/\psi$ in different spin states is then expected to be affected by the magnetic field.  On the other hand, in the Color-Octet Model (COM) for $J/\psi$ production~\cite{Bodwin:1994jh},  the charm and anticharm quark pair is first produced in a color octet state  and then converts to a $J/\psi$ by emitting a gluon  at a much later time when the strength of the magnetic field has significantly diminished.  As a result, the  spin orientation of produced $J/\psi$ in the COM is random and no spin asymmetry of  $J/\psi$ is  expected to be  produced in this case. Because it has been shown that both the $p_T$ spectrum and the polarization of $J/\psi$ in p+p collisions at the Tevatron energies can be reproduced by the upper limit of a higher order calculation in the CSM~\cite{Lansberg:2011hi}, we shall assume in the present study that $J/\psi$ production in heavy ion collisions at the LHC can be described by the CSM. By using the statistical hadronization model~\cite{Andronic:2008gu, BraunMunzinger:2000px} for these initially produced $J/\psi$, we then calculate the fraction $F$ of produced $J/\psi$s with $J_y=0$ to determine the spin asymmetry of $J/\psi$ in heavy ion collisions at the LHC. We remind the readers that the COM is an alternative model for $J/\psi$ production, and there is no spin asymmetry of produced $J/\psi$ in this case. Such a difference in the spin asymmetry of $J/\psi$ can be used to probe the production mechanism of $J/\psi$ in the initial hard collisions.

This paper is organized as follows. In Section~\ref{se_magnetic_field}, we give a simplified description of the magnetic field produced in heavy ion collisions. The motion in the phase space and the spin evolution of $J/\psi$s in the medium is described in Section~\ref{se_transport}, with the initial conditions introduced in Section~\ref{se_scm}.  We then show in Section~\ref{se_numerical} the numerical results  for the spin asymmetry of $J/\psi$ in heavy ion collisions at the LHC energy. Finally, a summary is given in Section~\ref{se_summary}.

\section{The magnetic field in relativistic heavy ion collisions}\label{se_magnetic_field}

Because most of the net charges pass through each other in peripheral collisions of heavy ions at relativistic energies,  all the nucleons can be essentially regarded as spectators.  The magnetic field in these collisions can be calculated by  superposing the Li\'enard-Weichert (LW) potentials generated by  the two colliding nuclei.  For two nuclei A and B with their centers at $x_A=b/2$ and $x_B=-b/2$, respectively, on the  $x$-axis at time $t=0$, where $b$  is the impact parameter of the collision,  and with velocity $v$\ parallel and anti-parallel, respectively,  to the unit vector ${\bf e}_z$ along the $z$-axis,  the magnetic field at ${\bf r}=(x,y,z)$ is
\begin{eqnarray}
   e{\bf B}^{\textrm{LW}}({\bf r}, t) &=&\gamma\alpha v \left(Z_A^*(r_{0A}){\bf e}_z\times\frac{{\bf r}-x_A{\bf e}_x}{r_{0A}^3}\right.\nonumber\\
   &-&\left.Z_B^*(r_{0B}){\bf e}_z\times\frac{{\bf r}-x_B{\bf e}_x}{r_{0B}^3}\right).
\end{eqnarray}
In the above, $e$ is the charge of a proton, $\alpha=e^2/(4\pi)$, and $Z^*_{A(B)}(r_{0})\equiv \int_0^{r_{0}}dr\ 4\pi r^2\rho_{eA(B)}(r)$\ is the effective charge inside  a volume of radius $r_{0}$ and charge density distribution $\rho_{eA(B)}(r)$, with $r_{0A}=\sqrt{(x-\frac{b}{2})^2+y^2+\gamma^2(z-v t)^2}$\ and $r_{0B}=\sqrt{(x+\frac{b}{2})^2+y^2+\gamma^2(z+vt)^2}$. In a peripheral collision, the magnetic field at the center of the collision is dominated by the $y$-component 
\begin{eqnarray}
   &&-eB_y^{\textrm{LW}}({\bf r}, t)\nonumber\\
   &=& -\gamma\alpha v \left(Z_A^*(r_{0A})\frac{x-b/2}{r_{0A}^3}-Z_B^*(r_{0B})\frac{x+b/2}{r_{0B}^3}\right).
   \label{eq_eby}
\end{eqnarray}

\begin{figure}[<!htb>]
   \begin{center}
	\includegraphics[width=0.47\textwidth]{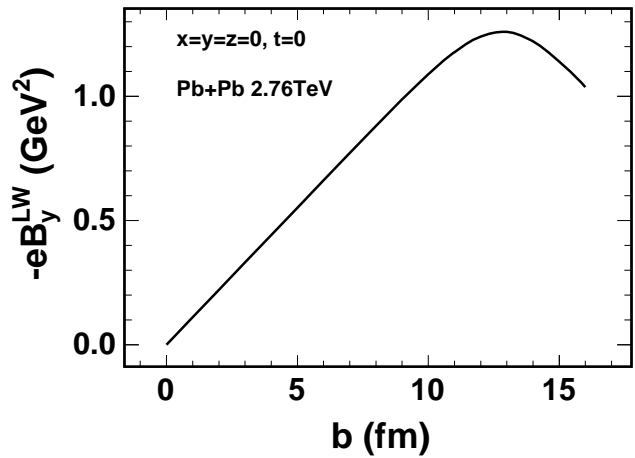}
   \end{center}
   \caption{Magnetic field $-eB_y^{\textrm{LW}}$\ in the direction perpendicular to the reaction plane at the middle of two Pb nuclei when they are side by side as a function of impact parameter $b$\ at the LHC energy of $\sqrt{s_{NN}}=2.76$ TeV.}
   \label{fig_eBy1}
\end{figure}

In the special case of A=B and $\bf r=\bf 0$, $t=0$, we obtain $-eB_y^{\textrm{LW}}({\bf 0}, 0) = 8\gamma\alpha v Z^*(b/2)/b^2$.  Its value as a function of $b$ in Pb+Pb collisions at the LHC energy of $\sqrt{s_{NN}}=2.76$ TeV is shown in Fig.~\ref{fig_eBy1}.  For $b/2$\  smaller than the radius $R$ of the nucleus, the ratio $Z^{*}(b/2)/(b/2)^3$\ is approximately $4\pi Z\rho_{0}/(3A)$, where $\rho_0=0.16/{\rm fm}^3$ is the  normal nuclear saturation density, and $Z$\ and $A$\ are the charge number and the mass number of the nucleus. Alternatively, we can write $-eB_y^{\textrm{LW}}({\bf 0}, 0)= k_s b$\ with $k_s= 4\pi Z\gamma v\alpha \rho_0/(3A)=0.11 \textrm{ GeV}^{2}/\textrm{fm}$ for Pb+Pb collisions at  $\sqrt{s_{\rm NN}} = 2.76$ TeV.  On the other hand, for $b/2$  larger than the radius $R$ , $Z^*(b/2)\approx Z$\ and the magnetic field is proportional to $1/b^2$. This explains well the behavior of $-eB_y^{\textrm{LW}}$\ in Fig.~\ref{fig_eBy1} and is consistent with that from the cascade simulation~\cite{Deng:2012pc}.

\par
The slope $k_s$\ above, as well as the maximum value of the magnetic field at LHC energy, is one order of magnitude larger than that at RHIC energy, mainly  because of the larger Lorentz contraction factor $\gamma$, so that the field is  localized in a smaller  region. Another consequence of the strong contraction is that the field decreases quickly. For $|t|\gg b/(2\gamma)$\ and $|t|\gg R/\gamma$, one finds $-eB_y^{\textrm{LW}}({\bf 0}, t) = \alpha v b Z/(\gamma^2|t|^3)$. Therefore, the large magnetic field decreases very fast at a later time of the collision as $eB_y^{\textrm{LW}}t^{-3}$.

\par
Because a large number of charged particles are produced in relativistic heavy ion collisions, the decreasing magnetic field can induce a current in the medium which in turn slows down the decrease of the magnetic field. However, the initial strong magnetic field still cannot be  sustained because of the fast expansion of the system in the longitudinal direction. In the large conductivity limit, the flux of magnetic field is constant, and  the magnetic field thus can be  approximately expressed as
\begin{eqnarray}
-eB_y({\bf x}, t)=
\begin{cases}
   -eB_y^{\textrm{LW}}({\bf x},0)\frac{L}{L+2v t}, &  t, L>0,\\
   -eB_y^{\textrm{LW}}({\bf x},t), &\textrm{otherwise},
\end{cases}
\label{eq_ebym}
\end{eqnarray}
where $L=L({\bf x}_T)$\ is the longitudinal thickness of the fireball at transverse coordinate ${\bf x}_T$\ and time $t=0$. In our calculations below, we take $L=\text{min}\left\{2\sqrt{R^2-({\bf x}_T-{\bf b}/2)^2}, 2\sqrt{R^2-({\bf x}_T+{\bf b}/2)^2}\right\}/\gamma$, and the radius $R=6.624$\ fm is determined from the  parametrized Woods-Saxon distribution of electric charges in  a Pb nucleus~\cite{De_Jager:1987qc}.

\begin{figure}[!hbt]
   \begin{center}
	\includegraphics[width=0.47\textwidth]{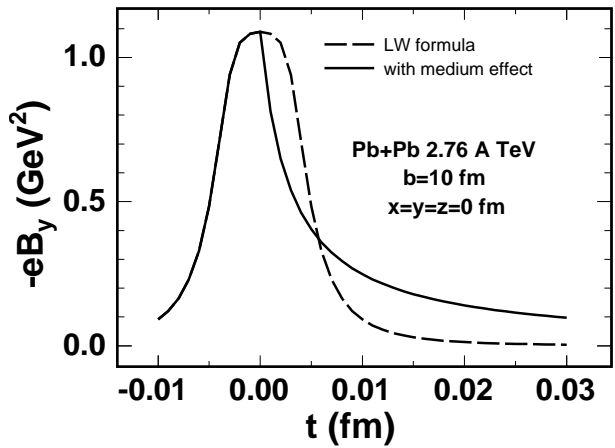}
   \end{center}
   \caption{Magnetic field in Pb+Pb collisions at  center of mass energy $\sqrt{s}=2.76$\ A TeV with impact parameter $b=10$\ fm at $x=y=z=0$\ fm. The dashed line is from the Lie\`nard-Weichert formula, see Eq.~(\ref{eq_eby}), and the solid line is with medium effects included, see Eq.~(\ref{eq_ebym}).}
   \label{fig_meBy}
\end{figure}

\par
In Fig.~\ref{fig_meBy}, we show the time evolution of the magnetic field in Pb+Pb collisions at the LHC energy. It can be seen that the free magnetic field $-eB_y^{\textrm{LW}}$ drops slowly at the very beginning because of the short distance from the center of the nuclei, and decreases fast later because it is proportional to  $t^{-3}$. With the inclusion of the strong medium effect,  the magnetic field $-eB_y$  decreases fast at the very beginning of the collision owing to the large expansion rate relative to its small initial volume, but  the decrease slows down as $-eB_y\propto 1/t$ at later times. In our following calculations, we use the  the magnetic field with the medium effect. Otherwise,  the effect of magnetic field on the spin asymmetry of $J/\psi$ is even stronger.

\section{The transport approach}\label{se_transport}

The motion of $J/\psi$\ in the medium is influenced by several effects. When the temperature is higher than the dissociation temperature $T_D$\ of $J/\psi$, the charm and anticharm quarks can no longer form a bound state  as a result of the screening effect.  We take the value of $T_D$ to be the same as that in Ref.~\cite{Liu:2012zw}, where the velocity dependence of $T_D$ due  to finite response time of the medium to the charm quarks inside $J/\psi$ is considered. For those surviving $J/\psi$s, their distribution function $f_{\psi}({\bf x}, {\bf p}, t)$ at coordinate ${\bf x}$, momentum ${\bf p}$\ and time $t$\ is described by the following transport equation:
\begin{eqnarray}
   (\partial_t + {\bf v}\cdot{\bf {\nabla}}) f_{\psi} &=&  - \alpha f_{\psi}-\gamma Mf_{\psi} + \beta,
	\label{eq_transport}
\end{eqnarray}
with $f_{\psi}=\left(f_{\psi}^{0}, f_{\psi}^{\pm}\right)^{T}$, where $f_{\psi}^0$\ and $f_{\psi}^{\pm}$\ are the distribution functions of $J/\psi$\ in the phase space with spin $J_y=0$\ and $|J_y|=1$, respectively. In the above, ${\bf v}={\bf p}/E_{\psi}$ is the velocity of $J/\psi$, and $M=\left(\begin{array}{cc}2&-1\\-2&1\end{array}\right)=\frac{3}{2}(1+\Sigma)$\ gives the relative flipping rate between $J/\psi(J_y=0)$\ and $J/\psi(|J_y|=1)$ with $\Sigma=\frac{1}{3}\left(\begin{array}{cc}1&-2\\-4&-1\end{array}\right)$\ being a square root of the identity matrix, and $\alpha$, $\beta$\ and $\gamma$\ are the dissociation rate, the regeneration rate and the spin flipping rate of $J/\psi$, respectively.  

The solution of Eq.~(\ref{eq_transport}) is 
\begin{eqnarray}
   f_\psi({\bf x}, {\bf p}, t)
   &=&K(t_0,t) f_\psi({\bf x}-{\bf v}(t-t_0), {\bf p}, t_0)\nonumber\\
   &+&\int_{t_0}^{t}K(\tau',t)\beta({\bf x}-{\bf v}(t-\tau'), {\bf p}, \tau')d\tau'.
   \label{eq_solution1}
\end{eqnarray}
In the above, $f_\psi({\bf x},{\bf p},t_0)$\ is the initial distribution at time $t=t_0$, and
\begin{eqnarray}
   K(t_0, t) &=& e^{-A(t_0,t)-G(t_0,t)}\nonumber\\
   &\times&\left(\cosh G(t_0,t)\right.- \left.\Sigma\sinh G(t_0,t)\right),
\end{eqnarray}
governs the time evolution with
\begin{eqnarray}
   A(t_0, t) &=&  \int_{t_0}^{t}\alpha({\bf x}-{\bf v}(t-\tau), {\bf p}, \tau)d\tau,\\
   G(t_0, t) &=&  \frac{3}{2}\int_{t_0}^{t}\gamma({\bf x}-{\bf v}(t-\tau), {\bf p}, \tau)d\tau
   \label{eq_k}
\end{eqnarray}
controling the dissociation of the charmonium and  the flipping of  its spin, respectively. Besides the time dependence shown explicitly, $K$, $A$, and $G$ all depend on ${\bf x}$\ and ${\bf p}$.  Although the ratio $\epsilon\equiv\gamma/\alpha$ of the average spin flipping cross section to the average dissociation cross section depends on details of both processes and thus  the temperature of the medium and the velocity of $J/\psi$,  we  take it as a constant for simplicity and show the results  for different values of $\epsilon$. From the above equations, we find that the spin flipping process leads to the following evolution of the fraction $f\equiv f_\psi^0/(f_\psi^0+f_\psi^{\pm})$ of a $J/\psi$: 
\begin{eqnarray}
   f(t)
   &=& \left(f(t_0)-\frac{1}{3}\right)e^{-3\epsilon A(t_0,t)}+\frac{1}{3}, 
   \label{eq_ft}
\end{eqnarray}
as long as it is not dissociated.

\par
 
For the $J/\psi$ dissociation rate, it can be expressed as
\begin{eqnarray}
\alpha({\bf x}, {\bf p}, t) &=&  
\frac{1}{E_{\psi}}\int\frac{d{\bf k}}{(2\pi)^3E_k}k_{\mu}p^{\mu}f_g^{\textrm{th}}\sigma.
\label{eq_alpha}
\end{eqnarray}
In the above, $p=(E_{\psi}, {\bf p})$\ and $k=(E_k, {\bf k})$\ are the momentum of the charmonium and the gluon, respectively, and $f_g^{\textrm{th}}(k, u, T)$ is the local thermal distribution of gluons with the local velocity $u^{\mu}$\ and  temperature $T$ determined by the hydrodynamic model as described in the last paragraph of this section. $\sigma(k, p)$\ is the cross section of the gluon dissociation process $J/\psi+g\rightarrow \bar{c}+c$, which is obtained by using the operator product expansion~\cite{Peskin:1979va, Bhanot:1979vb, Arleo:2001mp, Schroedter:2000ek}. 

\par 

Besides the dissociation of $J/\psi$, the inverse process $c+\bar{c}\rightarrow J/\psi+g$\ is included through the regeneration rate 
\begin{eqnarray}
   &&\beta({\bf x}, {\bf p}, \tau')\nonumber\\
   &=&\frac{1}{2E_{\psi}}\int\frac{d{\bf k}}{(2\pi)^32E_g}\frac{d{\bf q}_c}{(2\pi)^32E_c}\frac{d{\bf q}_{\bar c}}{(2\pi)^32E_{\bar c}}W(s)f_cf_{\bar{c}}\nonumber\\
   &\times&(1+f_g^{\textrm{th}})(2\pi)^4\delta^{(4)}(p+k-q_c-q_{\bar c})\left(\begin{array}{c}f(\tau')\\1-f(\tau')\end{array}\right).\nonumber\\
	\label{eq_beta}
\end{eqnarray}
In the above, $W(s)$\ is the transition probability, which is related to the dissociation cross section $\sigma$\ in Eq.~(\ref{eq_alpha}) by the detail balance relation, and $p$, $k$, $q_c$, $q_{\bar c}$ are momenta of the $J/\psi$, gluon, charm and anticharm quarks, respectively, and $f_c$, $f_{\bar c}$\ and $f_g^{\textrm{th}}$\ denote distribution functions of  corresponding particles. $f_g^{\textrm{th}}$\ is taken as thermal distribution as in the dissociation rate $\alpha$. For simplicity, we also assume kinetic thermalization for the heavy quarks, i.e. the momentum distribution is thermal while the density $\rho_c$\ respects the conservation law
\begin{eqnarray}
   \partial_{t}\rho_c+\nabla\cdot(\rho_c {\bf v}_c)&=&0,
\end{eqnarray}
where ${\bf v}_c$ is taken as the velocity of the medium ${\bf u}/u^0$. 

\par
The excited states are considered in the same way, but with different dissociation cross sections~\cite{Wang:2002ck, Arleo:2001mp} and lower dissociation temperatures as in Ref.~\cite{Liu:2012zw}.

\par
For the dynamics of heavy ion collisions, which is needed to describe  charmonium suppression and regeneration, we use a  2+1 dimensional ideal hydrodynamics with the assumption of boost invariance~\cite{Kolb:2000fha}.  The equations of state (EoS)  is taken as an ideal gas of hadrons and partons in the confined and deconfined phases, respectively, with a first order phase transition at $T_c=165$\ MeV~\cite{Aoki:2006br, Borsanyi:2010bp}. The initial condition is determined by the Glauber model, and the maximum temperature of the fireball in collisions at impact parameter $b=10.5$ fm is $353$ MeV at the thermalization time $t_0=0.6$ fm. The parameters for the fireball are the same as  those used in Ref.~\cite{Liu:2012zw}. Although this model is admittedly crude compared to currently available hydrodynamic models, it is useful for the present exploratory study of the spin asymmetry of $J/\psi$ in relativistic heavy ion collisions.

\section{the Initial distribution}\label{se_scm}

The initial conditions  for Eq.~(\ref{eq_transport}) include the unpolarized distribution $f_\psi(t_0)=f^{0}_{\psi}(t_0)+f^{\pm}_{\psi}(t_0)$ and the initial fraction $f(t_0)$ of $J/\psi$s with spin $J_y=0$. The former is obtained according to the Glauber model with the differential cross section for $J/\psi$\ in $pp$\ collisions parameterized as~\cite{Bossu:2011qe}
\begin{eqnarray}
   \frac{d^2\sigma}{dydp_T}&=&\left.\frac{d\sigma}{dy}\right|_{y=0}e^{-\frac{y^2}{2\xi^2y_{\textrm{m}}^2}}\frac{2(n-1)p_T}{D_y}  \left(1+\frac{p_T^2}{D_y}\right)^{-n}.
   \label{eq_differential_sigma}
\end{eqnarray}
In the above, $y$\ and $p_T$\ are the rapidity and transverse momentum of $J/\psi$, respectively. For the parameters, $\xi=0.39$\ is a constant, and $ y_{\textrm{m}}\equiv \ln(\sqrt{s_{NN}}/m_{J/\psi})$ is the maximum rapidity of a $J/\psi$\ with its mass $m_{J/\psi}$ produced at the beam energy $\sqrt{s_{NN}}$, with
\begin{eqnarray}
   D_y&\equiv& (n-2)\average{p_T^2}_y=\average{p_T}_y^2/a_n^2,\\
   a_n&\equiv&  (n-1)B(3/2,n-3/2),
\end{eqnarray}
where $B(x,y)$\ is the Beta function. The experimental results from the ALICE Collaboration implies $n=3.9$ and $\average{p_T} = 2.3$\ GeV~\cite{Bossu:2011qe} at forward rapidity $2.5<y<4$.  We further assume that the rapidity dependence of $\average{p_T^2}$\ satisfies
\begin{eqnarray}
   \average{p_T^2} (y) = \average{p_T^2} (y=0)\left(1-(y/y_{\textrm{m}})^2\right),
   \label{eq_pt2_y}
\end{eqnarray}
which well reproduces the results at RHIC~\cite{Adare:2006kf, Liu:2009wza}. We take $n=3.9$, $\langle p_T^2\rangle_{y=0}=9.0$\ GeV$^2$\ and $d\sigma/dy|_{y=0}=3.6\textrm{ }\mu$b according to above parameterization and the experimental results~\cite{Abelev:2012kr, Bossu:2011qe}. The Cronin effect is also considered in our calculation by smearing the momentum distribution from $pp$ collisions~\cite{Zhao:2007hh, Liu:2012zw}.

In the statistical hadronization model, the ratio of yields of different particles are determined by the thermal distribution, which successfully explains not only the experimental results in A+A collisions~\cite{Andronic:2008gu} but also that in  $pp$, $p\bar{p}$\ and $e^+e^-$\ collisions~\cite{Becattini:1996gy}. For heavy flavors production in $e^+e^-$\ collisions, the statistical model also describes the relative yields of different species of D-mesons and B-mesons very well~\cite{Redlich:2009xx, Andronic:2009sv}. We generalize this assumption to the spin degree of the constituent heavy quarks in both initially produced quarkonia from $p+p$\ collisions and regenerated quarkonia from the QGP.  The  relative yields of the four spin states of heavy quark pairs are thus
\begin{eqnarray}
   | \uparrow\uparrow \rangle : | \downarrow\downarrow \rangle : | \uparrow\downarrow \rangle : | \downarrow\uparrow \rangle = 1:1:e^{-E/T_{\textrm{eff}}}:e^{+E/T_{\textrm{eff}}},
   \label{eq_scm}
\end{eqnarray}
where $\ket{\uparrow}$\ and $\ket{\downarrow}$\ denote different spin states of  a heavy quark, and the energy difference between them is
\begin{equation}
   E = 2 \frac{q}{2m_c} B \approx \frac{2qB}{m_{J/\psi}} = \frac{2QeB}{m_{J/\psi}},
   \label{eq_energy_difference}
\end{equation}
In the above, $q=Qe$\ is the electric charge of a charm quark with $Q=2/3$, and $m_c$\ is the mass of  the charm quark. Although the temperature of the fireball would be high at the very beginning if it could be defined, neither thermal equilibrium nor chemical equilibrium between the heavy flavors and the medium is expected to be reached so early~\cite{vanHees:2004gq, Uphoff:2010sh, Cao:2011et, Das:2013kea}. The behavior of heavy flavors is thus dominated by that in elementary $p+p$ collisions. In our calculations, the  effective temperature $T_{\textrm{eff}}=0.17$\ GeV is taken as the same as that in $e^+e^-$\ collisions~\cite{Redlich:2009xx}\footnote{Based on PYTHIA ~\cite{Sjostrand:2006za, Sjostrand:2007gs} simulations for $p+p$\ collisions at $2.76$ TeV, the  effective temperature is estimated  to be also $0.17$ GeV for the yields of $K_0$\ and $K_0^*$ 
but is  somewhat smaller for those of $D$ and $D^*$ mesons, which would increase the effect of the magnetic field.}, which is also   similar to that in heavy ion collisions~\cite{Andronic:2008gu}. Note that the $J/\psi$\ with $J_y=0$\ is a mixed state of $\ket{\uparrow\downarrow}$\ and $\ket{\downarrow\uparrow}$, the yield of $J/\psi$s with different spins are related as $\ket{J_y=1}:\ket{J_y=-1}:\ket{J_y=0} = 1:1:\cosh(E/T_{\textrm{eff}})$.\footnote{The  duration of the magnetic field is short compared with the inverse energy change of $J/\psi$ due to the magnetic field. Therefore, the Hamiltonian  of the system changes suddenly, and both $\ket{\uparrow\downarrow}$\ and $\ket{\downarrow\uparrow}$ contribute to $J/\psi$, which is different from an adiabatic process.}  Thus the fraction of the $J_y=0$ component is
\begin{eqnarray}
   f &=& \frac{\cosh(E/T_{\textrm{eff}})}{2+\cosh(E/T_{\textrm{eff}})}.
   \label{eq_ratio}
\end{eqnarray}
In the absence of a magnetic field, i.e. $E=0$, we have $f=1/3$.  The maximum magnetic field $-eB_y=1.1$\ GeV$^2$\ at $b=10$\ 
fm shown in Fig.~\ref{fig_meBy} leads to an energy difference in Eq.(\ref{eq_energy_difference}) $E_{\textrm{max}}=0.5 
\textrm{ GeV}$ comparable to $T_{\textrm{eff}}$, which gives  an upper limit of the fraction $f_{\textrm{max}}=0.83$.  
For a charm pair produced from an initial nucleon-nucleon binary  collision, they are affected by the magnetic field only after a 
formation time $t_f$ when they are formed, which is much shorter than the formation time of $J/\psi$, which is  in the order of the inverse of its binding energy according to  the uncertainty relation.  
We assume that the formation time follows a Poisson distribution with the average value 
$\langle t_f\rangle=1/(2E_T)$ estimated by the uncertainty relation, where $E_T$\ is the transverse energy of the $J/\psi$. For 
regenerated $J/\psi$ from charm quarks  in the quark-gluon plasma, which is important for  low $p_T$ $J/\psi$ in heavy ion 
collisions at the LHC energy~\cite{Liu:2009nb, Zhao:2012gc},  the energy difference of $J/\psi$s with different spin states is 
negligible, because they are produced after the formation of the quark-gluon plasma at a time scale of $t\sim 1$\ fm~\cite{Liu:2009nba} when the magnetic field is several orders of magnitude smaller. The  regenerated $J/\psi$  thus have equal probabilities in 
its different spin states, and the fraction $f(\tau')$ in Eq.~(\ref{eq_beta}) is 1/3. 

\section{Results}\label{se_numerical}

\begin{figure}[!hbt]
   \begin{center}
	\includegraphics[width=0.47\textwidth]{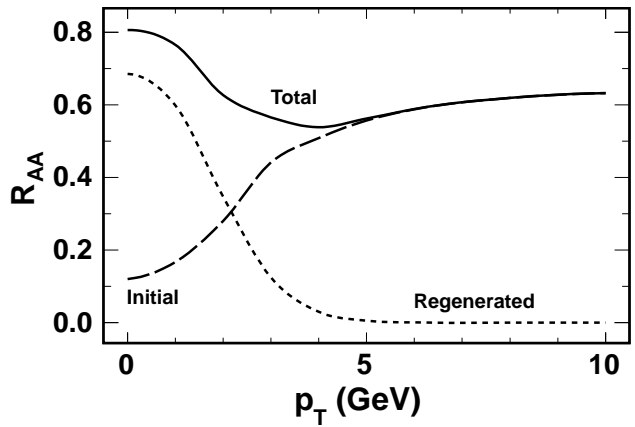}
   \end{center}
   \caption{The nuclear modification factor $R_{AA}$\ of prompt $J/\psi$\ at impact parameter $b=10.5$\ fm in Pb+Pb collisions at $\sqrt{s}=2.76$\ A TeV. The dashed, dotted, and solid lines are the contribution from the initially produced $J/\psi$s, the regenerated $J/\psi$s, and both, respectively.}
   \label{fig_raa}
\end{figure}

Before discussing the spin structure of $J/\psi$, we first show in Fig.~\ref{fig_raa}  the transverse momentum dependence of  its nuclear modification factor $R_{AA}(p_T)$ in peripheral collisions  at impact parameter $b=10.5$ fm. Because the suppression of initial $J/\psi$s with high transverse momentum is weak~\cite{Adamczyk:2012ey, Liu:2012zw}, the suppression at high $p_T$ is mainly due  to the excited states  that can hardly survive. As a result,  the $R_{AA}$\ at high $p_T$ is only slightly larger than the fraction $0.6$~\cite{Wang:2002ck} of direct $J/\psi$s.  Because charm quarks interacts strongly with the medium, which is indicated by the fact that the flow of $D$-mesons is comparable to that of light-flavor hadrons~\cite{Abelev:2013lca}, we have assumed that they are  kinetically thermalized  and most likely to have relatively small transverse momentum. As a consequence, the regenerated $J/\psi$s, which are produced from these  small transverse momentum  charm quarks,  contribute only to the low $p_T$ region.  Due to the competition of the initial production and the regeneration, there appears a minimum at $p_T=4$\ GeV. By integrating the transverse momentum, the total $R_{AA}$\ is found to be $0.65$, which is consistent with recent experimental results.~\cite{Arsene:2013vp}.

\begin{figure}[!hbt]
   \begin{center}
	\includegraphics[width=0.47\textwidth]{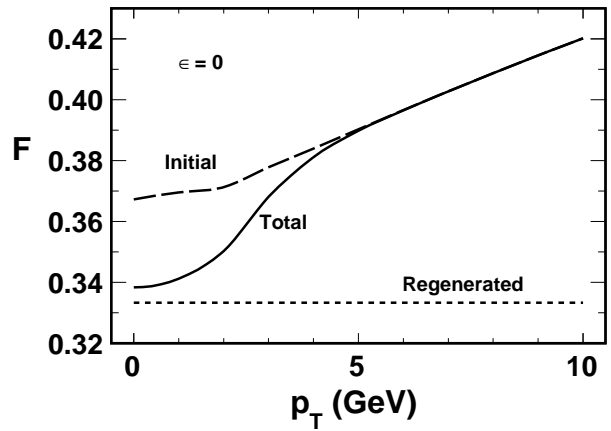}
   \end{center}
   \caption{The fraction  $F$  of $J/\psi$s with $J_y=0$\ without spin flipping in the medium ($\epsilon=0$) in peripheral $\textrm{Pb}+\textrm{Pb}$\ collisions at $b=10.5$\ fm at $\sqrt{s}=2.76$\ A TeV in the CSM scenario. The dashed, dotted, and solid lines are the fraction of the survived initially produced $J/\psi$s, the regenerated $J/\psi$s, and the total, respectively.}
   \label{fig_frac}
\end{figure}

The  ${\bf x}$\ and ${\bf p}_T$\ integrated fraction $F\equiv N_{J_y=0}/N$ of $J/\psi$ in the final state with no spin flipping in QGP ($\epsilon=0$) is shown in Fig.~\ref{fig_frac}, where $N_{J_y=0}$\ and $N$\ are the yield of $J/\psi$\ with $J_y=0$\ and the total yield, respectively.  Because 
the formation time of charm quark pairs $\average{t_f}=1/(2E_T)$ is $0.03$\ fm at $p_T=0$, which is large compared with the 
time for the magnetic field to decay (see Fig.~\ref{fig_eBy1}),  only a small part of the charm pairs feels the magnetic field. For 
those $J/\psi$s  of $p_T=10$ GeV, the average formation time of the charm pairs $\average{t_f}=0.01$ fm is much smaller,  
and more charm pairs  would have spin asymmetry.  The fraction $F$\ for initially produced $J/\psi$\ thus increases with transverse momentum. The regenerated $J/\psi$s are produced much later and  thus are not affected by the magnetic field, leading to $F=1/3$. Because the regeneration contributes the most to the population of finally observed $J/\psi$s at low $p_T$, as shown in Fig.~\ref{fig_raa}, the fraction $F$ is near  $1/3$ at $p_T\approx 0$. On the other hand,  the $F$ at high $p_T$ is essentially the same as that of the initially produced ones because their dominance in $J/\psi$ production at $p_T>5$ GeV. As a result, the fraction $F$ increases monotonously with $p_T$\ from $0.34$\ to $0.42$.

\par

\begin{figure}[!hbt]
   \begin{center}
	\includegraphics[width=0.47\textwidth]{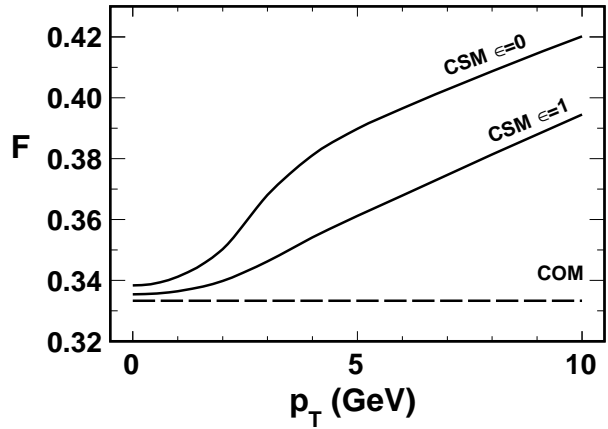}
   \end{center}
   \caption{The fraction  $F$  of $J/\psi$s with $J_y=0$\ in peripheral $\textrm{Pb}+\textrm{Pb}$\ collisions at $b=10.5$\ fm at $\sqrt{s}=2.76$\ A TeV\  for different parameter $\epsilon$\ for spin flipping in the CSM scenario (solid) compared with that in the COM scenario (dashed).} 
   \label{fig_epsilon}
\end{figure}

Besides the inelastic scattering  of $J/\psi$s in the QGP, there are also the elastic scattering  that can flip their spins. In Fig.~\ref{fig_epsilon}, we show the final fraction $F$  for different spin transition coefficient $\epsilon$ that was introduced in Section~\ref{se_transport}  to denote the ratio of the average flipping cross section to the average inelastic cross section. The elastic collision of the $J/\psi$  with the partons in QGP is usually considered as negligible in most of the models~\cite{Grandchamp:2002wp, Zhu:2004nw, Song:2010ix}, which is verified by the small elliptic flow of $J/\psi$\ at RHIC~\cite{Adamczyk:2012pw}. In this case, there is no spin transition, that is $\epsilon=0$\ as already shown in Fig.~\ref{fig_frac}. It is obvious from Eq.(\ref{eq_ft}) that the random flipping makes the fraction approach 1/3. For $\epsilon=0$, there is no flipping, and the fraction $F$\ can be as large as 0.42 around 10 GeV. Even if the flipping cross section is the same as the dissociation cross section, i.e. $\epsilon=1$, the fraction $F$\ can  still be above $0.39$\ at $p_T=10$ GeV, which is obviously larger than 1/3. For a more realistic value of $\epsilon$, which would be strongly model dependent, the spin fraction is expected to lie between these two extreme cases. We recall that the above discussion is based on the CSM scenario. If  $J/\psi$\ production is  described by the COM,  the $J/\psi$s mainly come from the spin singlet state of initially produced charm and anticharm quark pairs after the emission of a gluon~\cite{Wong:1999ur}. In this case, no $J/\psi$ spin asymmetry is expected, and the fraction is  simply 1/3 as shown by the dashed line in Fig.~\ref{fig_epsilon}.  Our results thus suggest that the measurement of the spin asymmetry of $J/\psi$ in heavy ion collisions can help  understand  its production mechanism  in initial hard collisions.

\section{Conclusions and outlook}\label{se_summary}

By generalizing the statistical hadronization model to the spin degree of freedom, we have calculated the influence of the magnetic field existing in the early stage of peripheral heavy ion collisions on the spin asymmetry of  produced $J/\psi$ at the LHC energy. The fraction of $J/\psi$s with spin $J_y=0$  is found to be above $1/3$ and to increase in peripheral Pb+Pb collisions with the transverse momentum in the CSM scenario, while it is $1/3$ in the COM scenario. Our finding thus indicate that studying the spin asymmetry of $J/\psi$ produced in relativistic heavy ion collisions provides the information of how the $J/\psi$ is produced in the initial hard collisions.  The present work can be generalized to study the asymmetry of dileptons produced in the initial stage of relativistic heavy ion collisions. 

\section*{Acknowledgements}

Liu is grateful to Xu-guang Huang for helpful discussions. This work was supported by the U.S. National Science Foundation under Grant No. PHY-1068572, the US Department of Energy under Contract No. DE-FG02-10ER41682, the Welch Foundation under Grant No. A-1358, and the Helmholtz International Center for FAIR within the framework of the LOEWE
program launched by the State of Hesse of Germany.

\bibliography{ref}

\end{document}